# Fluidic vortices generated from optical vortices in a microdroplet cavity

Daniel Bar-David,* Shai Maayani, Leoplodo L. Martin, and Tal Carmon

*Mechanical Engineering Department, Technion-Israel Institute of Technology, 32000 Haifa, Israel*
*Corresponding author: danielba@technion.ac.il*

**We harness the momentum of light resonating inside a micro-droplet cavity, to experimentally generate microflows within the envelope of the drop. We 3D map these optically induced flows by using fluorescent nanoparticles; which reveals circular micro-streams. The flows are parametrically studied and, as expected, exhibit an increase of rotation speed with optical power. The flow is non-circular only when we intentionally break the axial symmetry of the droplet. Besides the fundamental interest in light-flow interactions including in optofluidic cavities, the optically controlled flows can serve in bringing analytes into the maximum-power region of the microcavity.**

Optofluidic resonators [1-4] are useful in detecting proteins [5], nanoparticles [6, 7] and viruses [8]. Such microfluidic cavities can be made in the form of a submerge solid-resonator [9], a solid resonator with liquid inside[10-12], or in the form of a droplet resonator [13-19]. A droplet is a small volume of self-contained liquid that is bounded almost completely by free surfaces. Its liquid-phase boundary typically self-forms to its final shape governed by interfacial tension. The droplet benefits from a nearly atomic-scale surface smoothness, which is necessary for reducing optical scattering-losses. Additionally, unlike solid devices, the shape of the drop is described by a well-known equation. As an example, knowing the contact angle between the droplet and the stem holding it, allowed to calculate the shape of the droplet resonator [20]. A droplet can contain three types of resonances: optical, acoustical, and capillary. Each of these 3 waves has its special velocity and absorption coefficient; which can provide, altogether, access to six measurables. Among these measurables are viscosity, compressibility and surface tension, which are challenging to interrogate with the optical mode only. Another advantage of the droplet is that its modes are almost entirely contained within the liquid. As for these 3 droplet modes (optical, acoustical and capillary), the optical modes of a droplet were recently mapped [20], and the droplets acoustical[21] and capillary[22] modes where optically excited. Furthermore, Brownian oscillations in the droplet were optically interrogated [23].

Last and most relevant here, nanoparticles at the evanescent region of a submerged cavity were recently observed orbiting [24], locked to the evanescent tail of the mode. Such a whispering gallery mode (WGM) carousel [24] raises the question of whether optical momentum can similarly generate circulating flows in droplets. As a rough example, let's assume that a one microWatt laser is coupled into the circulating WGM of a droplet. Let's also assume that most of the optical losses in this resonator are due to absorption. Momentum conservation consideration implies that this absorbed microWatt must apply azimuthal force on the liquid that is tangential to its equatorial line.

Also relevant here is that in contrast to solids, where one generally has to overcome the static friction to start motion; liquid viscosity is typically speed independent. As a result, optically induced flows do not have an optical power threshold. Thus, there is no question if light will initiate flow; the only question is what will be the speed of this flow. Surprisingly maybe, the optically induced flows in the microdroplet cavity are independent of its optical quality factor. This is because it does not matter if the microWatt of optical power will be absorbed after circulating one round-trip or 1000 round-trips in the cavity. For in both cases a similar amount of momentum will be transferred to the microdroplet.

As for mapping the flows, we do so by inserting and mixing fluorescent nanoparticles in the liquid. As explained in section 5, we took special care in verifying that the nanoparticles are probing the flows while drifting at the exact speed of the liquid molecules. Although we will focus here on droplet resonators, the transfer of momentum from light to liquid as well as the resulting flows are relevant for any optofluidic device.

Assuming only absorption optical losses in the cavity, the optical force that pushes the liquid near the air-liquid interface to circulate is

$$F_0 = \frac{n \cdot P}{c} \quad (1)$$

where P is the input optical power, c is the speed of light and n the refractive index of the droplet liquid. The drag forces, Fs, which resist the optical forces, is estimated using the shear-flow equation:

$$F_s = \mu A \frac{u}{y} \quad (2)$$

where μ is the dynamic viscosity, u is to the speed of liquid at the air-liquid interface and y is the distance from the liquid-air interface to the stem at the region of motion. A is the area of the moving liquid at the liquid-air interface, and is equal to the circumference of the droplet multiplied by the width of the optical mode along the stem. As mapped in [20], the width of the mode is typically 20% of the droplet extension along the stem.

After accelerating and reaching steady state, the optical force (Eq. 1) is fully balanced by the drag force (Eq. 2). Using this relation, we can calculate the speed at the liquid-phase boundary:

$$u = \frac{yn}{\mu A c} P \quad (3)$$

To give a relevant scale, a typical droplet might be 100 µm in diameter and the stem holding it might be 10 µm in diameter. The liquid can have a viscosity of 10 cSt (10 times higher than water) and a refractive index equal 1.399. Coupling 4 microwatt of light to such a droplet will induce a 9.22 µm/s flow-speed according to equation 3. This estimation in fact implies that flows exist in almost any optofluidic resonator including the ones described in [4-24].

We fabricate our droplet cavities by dipping a tapered silica fiber in silicone oils (Sigma-Aldrich #378321 or #378356) [20, 21]. As a result of interfacial tension between the solid anchor and the liquid, droplets tend to drift toward the thicker part of the taper to maximize the solid-liquid interface. We prevent this drift and pin the droplet to the taper end by turning the solid taper-end into a small sphere, using a fiber splicer.

As one can see in Figure 1, we use a tapered fiber to evanescently couple [25-27] a tunable laser (970 nm) into the droplet resonator. The light transmitted through the droplet is coupled out through the other side of the taper and measured with a photodetector. A nano positioning system is used to control the distance between the coupler and the resonator.

This optically-pumped droplet per-se is sufficient for optically induced flows. However, in order to film these flows, we mix fluorescent polymer nanospheres (Bang FS02F envy green) in the liquid. We film the nanoparticle's path which represents the flow streamline. Furthermore, when defocused, the microscope smears the nanoparticle's image into a larger spot. The size of the yellow fluorescent spot, together with its diameter, allows us therefore to get its position in 3 dimensions.

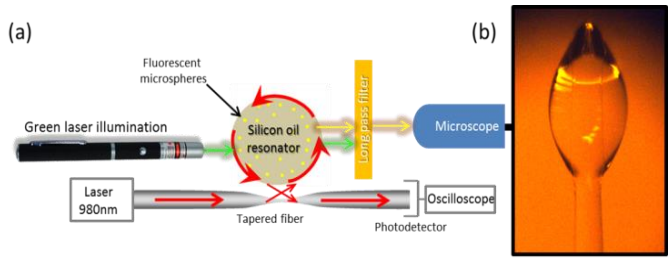

Fig. 1. Experimental setup. (a) The light is coupled via a tapered fiber to an optical WGM of a droplet resonator. Fluorescent microspheres are excited by a green laser pointer that the fluorescent nanoparticles convert to yellow. (b) A micrograph of a silica micro stem holding a silicon oil droplet through a yellow pass filter, before the nanoparticles are fluorescently excited.

For minimizing the optical forces on the nanoparticles probe, we use small nanoparticles with low absorption coefficient. The mean diameter of the nanospheres is 60 nm and theirs refractive index is 1.57. The nanospheres are fluorescently excited by a green laser (532nm) and emit yellow light (565nm). Using a yellow-pass filter, we film the streamlines by tracking the nanoparticles.

In what follows, we monitor the optically induced flows while changing the optical input power, the resonator size and the viscosity of the liquid.

As one can see in figure 2 (violet), increasing the input power from 0 to 4 µW, for a 10cst-viscosity droplet with a diameter of 100 µm and a refractive index of 1.399 reveals a linear increase of the velocity from rest to 12.5 µm/s. Such a linear behavior is expected from equation 3.

We now repeat this experiment while changing the diameter of the droplet to 200 µm. As one can see in figure 2 (green) speed goes down with resonator size.

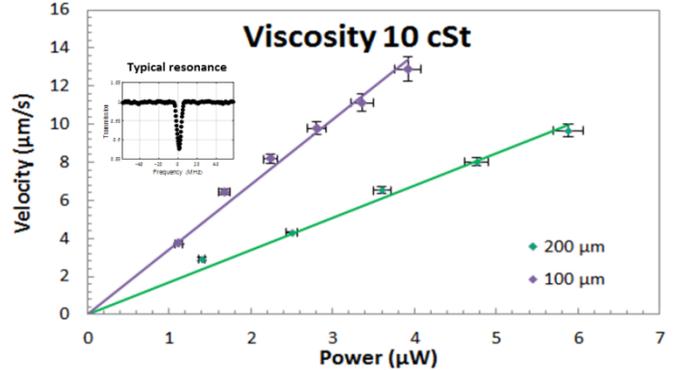

Fig. 2. Flow speed as a function of the input optical power in droplets 10 cSt in kinematic viscosity and different diameters of 100µm and 200µm. The violet (green) curve is the fitted linear line to the measured speeds in the 100 µm (200 µm) with an $R^2$ =0.978 (0.981). The flow velocity was calculated by averaging three measurements and the error bar of the velocity is the standard deviation.

Using equation 3, we can calculate that a rise of a droplet radius from 100 µm to 200 µm with the same 10cSt viscosity, and assuming that the mode area in the bigger droplet is twice as large as the mode area in the smaller droplet, will reduce the speed by 2 (while the experimental speed reduction is 1.97).

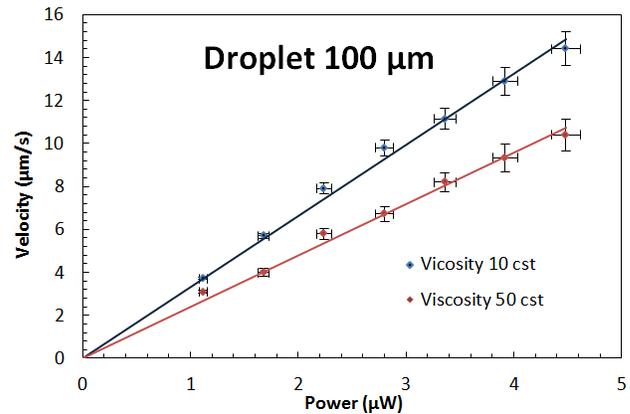

Fig. 3. Flow speed as a function of optical power for droplets of two different viscosities. The blue (red) curve represents a linear fit to the measured velocity for a droplet having a 10 cSt (50 cSt) viscosity with an $R^2$ = 0.992 (0.988). The error bar was calculated as explained in figure 2 caption.

We then repeat the experiment measuring the speed as a function of an input power, but this time at a higher viscosity which slows down the flow. As expected from equation 3, increasing the viscosity indeed reduced the speed; yet, the exact amount of speed reduction is less that what is predicted by equation 3. One explanation for that might be that the high viscosity liquid has more scattering losses when compared to the low viscosity liquid.

As one can see in figure 4 a, moving the particle toward the camera results in blurring. By calibrating this blur diameter, we can estimate the distance of the particle from the focal plan where its image has the smallest diameter. In order to get a lookup table of the distance to the microscope as a function of spot size, we take a coverslip, put on a fluorescent nanoparticle and captured images with difference focus by controlling the distance with a XYZ piezo stage. This measured relation between blurring and distance is presented in Fig 4 a.

We merge 18 frames from a movie of the flowing nanoparticle and present them in figure 4b. As one can see (Fig 4b), the position and bluer of the 18 spots reveal the 3D position of the nano particle for each of the 18 measurements.

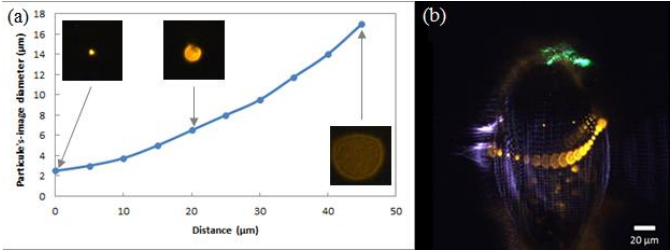

Fig. 4. (a) Calibration from a control group: the blue curve shows the fluorescent nanoparticle image diameter as a function of the focus distance (b) Merging of video frames showing the path of a nanosphere during half round in a droplet with a diameter of 100 μm presented in Media 1.

Using this method, we now map the nanoparticle trajectory in 3D. As one can see in Figure 5, the streams are almost circular.

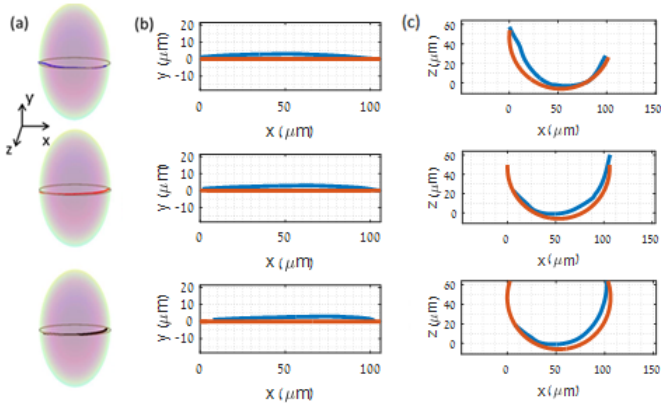

Fig. 5. (a) Fluorescent particles path of three successive rounds in a silicone oil droplet with a diameter of 100μm form the video of Fig 5(b). Each round is presented in a different color: blue, red and black. (b) XY plane for each rotation. (c) XZ plane for each rotation. The blue curve is the path of each lap.

In Fig. 6 we present instabilities observed in a droplet with a diameter of 250μm where due to the gravity, there is a fall in the droplet symmetry and the flow is non circular. The anchor of this droplet is 28 μm away from the center and the deviation of the droplet from a circular shape is 3%.

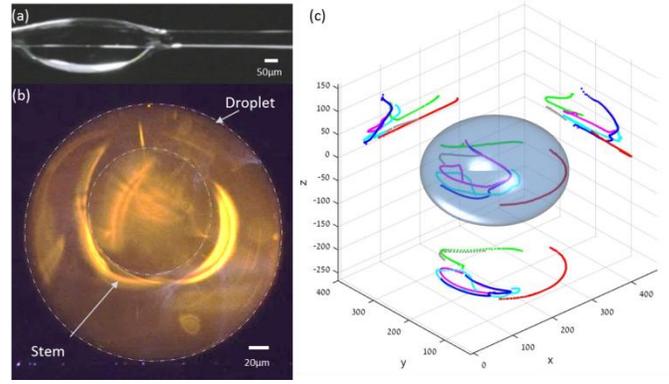

Fig. 6. (a) A micrograph of a silica micro stem holding a non-axially symmetric silicon oil droplet (b) Video showing the path of fluorescent nanospheres in a non-axially symmetric resonator presented in Media 2 (c) Three dimension representation of droplet instabilities with a large diameter of 250μm. Each continuous line represents the path of a nanosphere and accompanied by its projections on the XY, XZ and YZ planes.

To conclude our measurements, we were examining resonators with axial symmetry as well as resonators which are non axially symmetric which were accordingly hosting circular and non-circular flows. The flow speed increases with power and decreases with resonator size and with viscosity.

As the nanoparticle functions here as a probe, it is important to show that the nanoparticle drifts with flow. In what follows, we will show that the nanoparticle does not move with respect to the flow. We will start by calculating the force that light applies on the nanoparticle.

Using the Rayleigh scattering equation, one can calculate the cross section area for scattering, $\sigma_{sca}$, using [28]:

$$\sigma_{sca} = \frac{2\lambda^2}{3\pi}\alpha^6 \left|\frac{m^2-1}{m^2+2}\right|^2 \tag{4}$$

where $\alpha = \frac{2\pi r}{\lambda}$, $m = \frac{n}{n_1}$, $\lambda$ is the incident wavelength, $r$ is the radius of the sphere and $n$ and $n_1$ are the refractive index of the sphere and the silicone oil respectively.

Throughout the experiment the nanospheres are suspended in the silicon oil droplet, the particles within the optical mode volume will experience the strongest scattering forces.

The intensity within the mode volume is:

$$I = \frac{F \cdot P}{A_m} \tag{5}$$

where F is the optical finesse, P is the source power and $A_m$ is the mode area in a direction perpendicular to propagation.

Taking the most pesimistic assumption that all scattering is in the backward direction we get the strongest force that light can apply on the nanoparticle:

$$F_{scat} = \frac{2 \cdot \sigma_{sca} \cdot I \cdot n_1}{c} \quad (6)$$

where c is the light speed.

Now if the particle is moving in respect to the liquid, it will experience a drag force:

$$F_d = 6\pi\mu r v \quad (7)$$

where v is the particle speed relative to the surrounding medium, $\mu$ is the dynamic viscosity.

At steady state, the drag force (Eq.7) is equal to the scattering force (Eq.6), which gives us the nanoparticle-liquid relative speed:

$$v = \frac{\sigma_{scat} \cdot F \cdot P \cdot n_1}{3 \cdot \pi \cdot \mu \cdot c \cdot r \cdot A_m} \quad (8)$$

From Eq. 8, we find out that in our experimental conditions, the fastest nanoparticle-liquid speed is 0.142 μm/s which is 102 times slower than the flow speeds measured on the liquid. One can therefore say that the nanoparticle motion represent the flow motion with an accuracy better than 1%.

An additional source of error might be the Brownian motion of the nano particle. We were therefore performing a control-group experiment where the nanoparticles were monitored while the pump was off. The nanoparticle drift during the time of a typical measurement (such as in Figure 4-6) was 3.1 μm, which is less than the thickness of the line in figures 5-6

We experimentally map streamlines in an optofluidic whispering-gallery resonator, and reveal that optically induced vortices are abundant in the fluidic whispering-gallery embodiment. Our work is relevant to optofluidic devices including microcavities that part of their optical mode overlaps with liquids.

**Funding**. This research was supported by ICore: the Israeli Excellence center 'Circle of Light' grant no. 1902/12 and by the Israeli Science Foundation grant no. 2013/15.